\documentclass[a4paper,5p,sort&compress]{elsarticle}

\usepackage{xcolor}
\usepackage{mathtools}
\usepackage{slashed}
\usepackage[colorlinks]{hyperref}
\usepackage[title,titletoc,toc]{appendix}
            
\usepackage[utf8]{inputenc}
            
\allowdisplaybreaks[0]

\makeatletter
\def\ps@pprintTitle{%
 \let\@oddhead\@empty
 \let\@evenhead\@empty
 \def\@oddfoot{}%
 \let\@evenfoot\@oddfoot}
\makeatother

\newcommand{\Zp}{{Z^\prime}}
\newcommand{\mZp}{m_{Z^\prime}}

\newcommand{\GeV}{{\,\rm GeV}}

\newcommand{\fb}{{\, \rm fb}}

\newcommand{\mO}{{\mathcal O}}
\newcommand{\mC}{{\mathcal C}}

\newcommand{\nn}{\nonumber}

\begin{document}

\begin{frontmatter}
  
\title{\bf \boldmath Correlating the CDF $W$-boson mass shift with the $b \to s \ell^+ \ell^-$ anomalies}
  
\author[ccnu]{Xin-Qiang Li}
\ead{xqli@mail.ccnu.edu.cn}

\author[ccnu]{Ze-Jun Xie}
\ead{xiezejun@mails.ccnu.edu.cn}

\author[ccnu,zzu]{Ya-Dong Yang}
\ead{yangyd@ccnu.edu.cn}

\author[ccnu]{Xing-Bo Yuan}
\ead{y@ccnu.edu.cn}

\address[ccnu]{Institute of Particle Physics and Key Laboratory of Quark and Lepton Physics~(MOE),\\
  Central China Normal University, Wuhan, Hubei 430079, China}

\address[zzu]{School of Physics and Microelectronics, Zhengzhou University, Zhengzhou, Henan 450001, China}

\begin{abstract}
The recently updated measurement of the $W$-boson mass by the CDF collaboration exhibits a $7\sigma$ deviation from the SM expectation, which may imply a sign of new physics beyond the SM. The observed discrepancy could be explained by new fermions that carry the electroweak gauge charges and affect the vacuum polarization of gauge bosons. Notably, if the new fermions also have the same quantum numbers as of the SM quarks, they can mix with the latter and thus modify the penguin diagrams governing the $b \to s \ell^+ \ell^-$ transitions. Therefore, the $W$-boson mass shift could be related to the $b \to s \ell^+ \ell^-$ anomalies observed by the LHCb collaboration during the past few years. To investigate this possibility, we consider in this paper a model containing a vector-like top partner gauged under a new $U(1)^\prime$ symmetry. It is found that the latest CDF $m_W$ measurement and the $b \to s \ell^+ \ell^-$ anomalies can be simultaneously accommodated at $2\sigma$ level.  
\end{abstract}

\end{frontmatter}

\section{Introduction}
\label{sec:introduction}

The Standard Model (SM) of particle physics has been remarkably successful in explaining most phenomena observed in experiments~\cite{Zyla:2020zbs}. After the discovery of the Higgs boson~\cite{ATLAS:2012yve,CMS:2012qbp}, one important goal of the Large Hadron Collider (LHC) and its high-luminosity upgrade is the direct searches for new phenomena beyond the SM~\cite{ZurbanoFernandez:2020cco}. While the LHC has not yet discovered any direct evidence of new physics (NP) beyond the SM, several interesting deviations from the SM predictions have been emerging from the precision measurements, such as the anomalies observed in $b \to s \ell^+ \ell^-$ processes~\cite{Bifani:2018zmi,Albrecht:2021tul,London:2021lfn}.

In the field of precision measurements, the electroweak (EW) precision observables have played an important role in establishing the structure of the EW sector of the SM and can provide a sensitive probe of NP~\cite{Hollik:1988ii,Langacker:1991zr,Erler:2019hds}. Recently, using the complete dataset collected by the CDF II detector at the Fermilab Tevatron, corresponding to $8.8\,\fb^{-1}$ of integrated luminosity, the CDF collaboration has reported an updated measurement of the $W$-boson mass~\cite{CDF:2022hxs}
\begin{align}
  m_W^{\rm CDF} = 80.4335 \pm 0.0064_{\rm stat} \pm 0.0069_{\rm syst}\GeV.
\end{align}
Interestingly, such a high precision measurement shows a $7\sigma$ deviation from the SM expectation obtained by a global EW fit, $m_W^{\rm SM}=80.357 \pm 0.006\GeV$~\cite{Zyla:2020zbs}. The latest CDF result also shows a significant shift compared to the average of the previous measurements from LEP, CDF, D0 and ATLAS, $m_W^{\rm PDG}=80.379\pm 0.012\GeV$~\cite{Zyla:2020zbs}, as well as the LHCb measurement, $m_W^{\rm LHCb} = 80.354 \pm 0.031\GeV$~\cite{LHCb:2021bjt}. If confirmed by future measurements, the discrepancy could imply a sign of NP beyond the SM and has motivated numerous studies of the phenomenological implications of such a marvellous measurement (see, \textit{e.g.}, refs.~\cite{Strumia:2022qkt,Asadi:2022xiy,Gu:2022htv,Lu:2022bgw,deBlas:2022hdk,Fan:2022yly,Liu:2022jdq,Bagnaschi:2022whn,Paul:2022dds,Heo:2022dey,Zheng:2022irz,Almeida:2022lcs,Cheng:2022aau,Chen:2022ocr,Alguero:2022est,Zhou:2022cql,Popov:2022ldh,Ghorbani:2022vtv,Bhaskar:2022vgk,Heckman:2022the,DiLuzio:2022xns,Fan:2022dck,Tang:2022pxh,Borah:2022obi,Sakurai:2022hwh,Athron:2022qpo,Athron:2022isz}).

The latest CDF $W$-boson mass shift could be induced by new fermions, which carry the EW gauge charges and contribute to the vacuum polarization of gauge bosons~\cite{Lee:2022nqz,Crivellin:2022fdf,Endo:2022kiw,Balkin:2022glu,Cao:2022mif,Baek:2022agi}. Notably, if the new fermions also have the same quantum numbers as of the SM quarks, they can mix with the latter and thus modify the $\gamma$- and $Z$-penguin diagrams that govern the flavour-changing neutral-current (FCNC) $b\to s$ transitions. Therefore, the observed discrepancy of the latest CDF $m_W$ measurement could be related to the $b \to s \ell^+ \ell^-$ anomalies observed by the LHCb collaboration~\cite{LHCb:2017avl,LHCb:2021trn,LHCb:2021lvy,LHCb:2020lmf,LHCb:2020gog}. To investigate this possibility, we consider in this paper the NP model introduced in ref.~\cite{Kamenik:2017tnu}. Specifically, the model is characterized by a vector-like top partner with an additional $U(1)^\prime$ gauge symmetry, where the top quark and the top partner have the same quantum numbers under the SM gauge symmetry, while only the latter is charged under the $U(1)^\prime$ symmetry. Similar models have been investigated, \textit{e.g.}, in refs.~\cite{Fox:2011qd,Berger:2012ec,Greiner:2014qna,Cox:2015afa,Kim:2016plm,Fox:2018ldq,Crivellin:2020oup}.

This paper is organized as follows. In section~\ref{sec:Zp}, we introduce the NP model. In section~\ref{sec:EWPT}, the top-partner contributions to the $W$-boson mass shift and the oblique parameters $S$, $T$ and $U$ are investigated. In section~\ref{sec:b2s}, we consider and compute the NP contributions to the $b \to s \ell^+ \ell^-$ transitions. In section~\ref{sec:numerics}, we first derive the constraints on the model parameters from the global EW fit in view of the latest CDF $m_W$ measurement, and then investigate the possibility of simultaneously explaining the observed $W$-boson mass shift and the $b \to s \ell^+ \ell^-$ anomalies. Our conclusion are finally made in section~\ref{sec:conclusion}. For convenience, the loop functions involved are relegated in the appendix.  

\section{Model}
\label{sec:Zp}

As mentioned in the last section, to simultaneously accommodate the latest CDF $W$-boson mass shift and the $b \to s \ell^+ \ell^-$ anomalies, the new fermions contributing to the vacuum polarization of gauge bosons should have the same quantum numbers as of the SM quarks. To this end, one simple realization is the $\Zp$ model introduced in ref.~\cite{Kamenik:2017tnu}. The model is characterized by a new $U(1)^\prime$ gauge symmetry that is spontaneously broken by the vacuum expectation value (vev) of a new scalar field $\Phi$, transforming as $(1, 1, 0, q_t)$ under the $SU(3)_C\otimes SU(2)_L \otimes U(1)_Y \otimes U(1)^\prime$ gauge symmetry. While all the SM fields do not carry $U(1)^\prime$ charge, a coloured singlet vector-like top partner $U_{L,R}^\prime$, transforming as $(3,1,2/3,q_t)$, is introduced in the model. Specific to the case where only the third-generation SM quarks mix with the top partner, the general renormalizable interactions take the form~\cite{Kamenik:2017tnu,Fox:2018ldq}
\begin{align}\label{eq:Lint}
  \mathcal{L}_{\text {int }}= &\big( \lambda_{H} \bar{Q}_{3 L} \tilde{H} u_{3 R}+\lambda_\Phi \bar{U}_{L}^{\prime} u_{3 R} \Phi +\mu \bar{U}_{L}^{\prime} U_{R}^{\prime}+\text {h.c.} \big)
  \nonumber \\[0.15cm]
  &\qquad +q_t g_{t} \left(\bar{U}_{L}^{\prime} \gamma^{\mu} U_{L}^{\prime}+\bar{U}_{R}^{\prime} \gamma^{\mu} U_{R}^{\prime}\right) Z_{\mu}^{\prime},
\end{align}
where $q_t$ and $g_t$ denote the charge and coupling of the $U(1)^\prime$ gauge symmetry, respectively. $Q_{3L}=(u_{3L},d_{3L})^T$ and $u_{3R}$ denote the third-generation left-handed quark doublet and right-handed quark singlet in the SM, respectively. Matter fields in the above Lagrangian are all given in the interaction eigenbasis. For simplicity, the scalar potential and the interaction between $\Zp$ and $\Phi$ are suppressed here. 

After the $U(1)^\prime$ and the EW symmetry breaking, where the vevs of the two scalar fields are given respectively by $\langle\Phi\rangle=v_{\Phi}/\sqrt{2}$ and $\langle H\rangle=v_{H}/\sqrt{2}\simeq 174\GeV$, both the SM fermions and the top partner as well as the $\Zp$ boson obtain their masses. By diagonalizing the mass matrix, the physical top quarks $(t_L,t_R)$ and its partners $(T_L,T_R)$ can be expressed in terms of the fermion fields in eq.~\eqref{eq:Lint} through the following rotation matrices~\cite{Fox:2018ldq}:
\begin{align}
\begin{pmatrix}
t_{L} \\
T_{L}
\end{pmatrix}
  &=
\begin{pmatrix}
\cos \theta_{L} &-\sin \theta_{L} \\
\sin \theta_{L} & \hphantom{-}\cos \theta_{L}
\end{pmatrix}
\begin{pmatrix}
u_{3 L} \\
U_{L}^{\prime}
\end{pmatrix}
,\\[0.2cm]
\begin{pmatrix}
t_{R} \\
T_{R}
\end{pmatrix}
  &=
\begin{pmatrix}
\cos \theta_{R} & -\sin \theta_{R} \\
\sin \theta_{R} & \hphantom{-}\cos \theta_{R}
\end{pmatrix}
\begin{pmatrix}
u_{3 R} \\
U_{R}^{\prime}
\end{pmatrix}
,
\end{align}
where the mixing angles $\theta_L$ and $\theta_R$ parameterize the rotation matrices of the left- and right-handed quarks, respectively. In terms of the physical parameters, the top-quark mass $m_t$, the top-partner mass $m_T$, as well as the $\Zp$-boson mass $m_{\Zp}= g_{t} v_{\Phi}$, the two mixing angles are related to each other through
\begin{align}\label{eq:theta_L}
  \tan\theta_{L}=\frac{m_{t}}{m_{T}} \tan \theta_{R},
\end{align}
with $m_t$ and $m_T$ determined, respectively, by
\begin{align}
  \lambda_{H}&=\frac{\cos \theta_{L}}{\cos\theta_R} \frac{\sqrt{2}\, m_{t} }{v_H}, \nonumber \\
  \left|\lambda_\Phi\right|&= g_{t} \left|\cos \theta_{L}\sin \theta_{R} \right|
                                        \frac{\sqrt{2}\, m_{T}}{m_{Z^{\prime}}}\left(1- \frac{m_{t}^2}{ m_{T}^{2}}\right). 
\end{align}

Turning to the fermion mass eigenbasis, the explicit expressions of the gauge interactions involving the top quark and the top partner can be written as~\cite{Kamenik:2017tnu,Fox:2018ldq}
\begin{align}
  \mathcal L_\gamma &= \frac{2}{3} e \bar{t} \slashed{A} t+\frac{2}{3} e \bar{T} \slashed{A} T\,, \label{eq:int1} \\[0.2cm]
  \mathcal L_W &= \frac{g}{\sqrt{2}}V_{td_i} \left(c_L \bar{t}\,\slashed{W} P_{L} d_i+s_L \bar{T}\, \slashed{W} P_{L} d_i\right)+\text { h.c.}\,, \label{eq:int2} \\[0.2cm]
  \mathcal L_Z &= \frac{g}{c_W}\left(\bar{t}_{L}, \bar{T}_{L}\right)
\begin{pmatrix}
\frac{1}{2} c_L^2-\frac{2}{3} s_W^2 & \frac{1}{2} s_Lc_L \\[0.15cm]
\frac{1}{2} s_Lc_L & \frac{1}{2} s_L^2-\frac{2}{3} s_W^2
\end{pmatrix} \slashed{ Z}
\begin{pmatrix}
t_{L} \\[0.15cm]
T_{L}
\end{pmatrix}    \nonumber   \\[0.15cm]
& + \frac{g}{c_W}\left(\bar{t}_{R}, \bar{T}_{R}\right) \bigg(-\frac{2}{3}s_W^2 \bigg) \slashed{ Z}
\begin{pmatrix}
t_{R} \\
T_{R}
\end{pmatrix}\,, \label{eq:int3} \\[0.2cm]
\mathcal L_{\Zp} &= q_{t} g_{t}\left(\bar{t}_{L}, \bar{T}_{L}\right)
  \begin{pmatrix}
s_L^2 & -s_Lc_L \\[0.15cm]
-s_Lc_L & c_L^2
\end{pmatrix}
\slashed{Z}^{\prime}
\begin{pmatrix}
t_{L} \\[0.15cm]
T_{L}
\end{pmatrix} \nonumber \\[0.15cm]
  & + (L \to R)\,, \label{eq:int4}
\end{align}
where $s_{L,R}=\sin\theta_{L,R}$, $c_{L,R}=\cos\theta_{L,R}$, and $s_W=\sin\theta_W$ with the Weinberg mixing angle $\theta_W$; $d_i$ stand for the down-type quarks $d,s,b$, $g$ denotes the SM weak coupling constant, and $V_{td_i}$ are the CKM matrix elements. The Yukawa interactions of the physical SM Higgs $h$ with the top quark and the top partner are given by
\begin{equation}\label{eq:Yukawa}
\mathcal L_h=\frac{m_{t}}{v_{H}}\left(\bar{t}_{L}, \bar{T}_{L}\right) \left(\begin{array}{cc}
c_L & c_L \tan \theta_{R} \\[0.15cm]
s_L & s_L \tan \theta_{R}
\end{array}\right) h
\begin{pmatrix}
t_{R} \\[0.15cm]
T_{R}
\end{pmatrix}+\text{h.c.}
\end{equation}

Assuming that the effective $\Zp$ couplings to the charged leptons are flavour diagonal and focusing only on its interaction with the muon, we can then write the general renormalizable $\mu^+ \mu^-\Zp$ vertex as~\cite{Crivellin:2015era,Kamenik:2017tnu,Fox:2018ldq}
\begin{align}
\mathcal L_\mu=\bar{\mu} \slashed{Z}^{\prime}\left(g_{\mu}^{L} P_{L}+g_{\mu}^{R} P_{R}\right) \mu\,,
\end{align}
where the effective couplings $g_\mu^L$ and $g_\mu^R$ depend on the specific UV completions. For example, one can introduce vector-like leptonic partners to generate such effective couplings~\cite{AristizabalSierra:2015vqb}. In this paper, since we only concentrate on the possible connection between the latest CDF $W$-boson mass shift and the $b \to s \mu^+ \mu^-$ anomalies, details of the lepton sector are not so much relevant. Thus, as in refs.~\cite{Kamenik:2017tnu,Fox:2018ldq,Crivellin:2015era}, we shall consider the general effective $\mu^+\mu^-\Zp$ interaction and leave the complete UV completions for future work.  

\section{$W$-boson mass shift and oblique parameters}
\label{sec:EWPT}

The global fit of the SM to the EW precision data, known as the global EW fit~\cite{Hollik:1988ii,Langacker:1991zr,Erler:2019hds}, is a powerful tool to explore the validity of the SM and provides also a sensitive probe of NP beyond it~\cite{Flacher:2008zq,Baak:2014ora,Haller:2018nnx,deBlas:2021wap}. As the EW parameters of the SM are closely related to each other, it is generally expected that some observables in the global EW fit may be affected once the latest CDF $m_W$ measurement is considered~\cite{Strumia:2022qkt,Asadi:2022xiy,Gu:2022htv,Lu:2022bgw,deBlas:2022hdk}. Most of the NP effects on the EW sector can be parameterized in terms of the oblique parameters $S$, $T$ and $U$~\cite{Peskin:1991sw,Peskin:1990zt,Maksymyk:1993zm}. These parameters are in turn related to the NP contributions to the vacuum polarization of gauge bosons and can be written, respectively, as~\cite{Peskin:1991sw,Peskin:1990zt}
\begin{align}
  S &= \frac{4 s_{W}^{2}c_{W}^{2}}{\alpha} \bigg[\frac{\Pi_{Z Z}\left(m_{Z}^{2}\right)- \Pi_{Z Z}(0)}{m_{Z}^{2}} \nn \\
     &\qquad\qquad \qquad -\frac{c_{W}^{2}-s_{W}^{2}}{s_{W} c_{W}}  \Pi_{Z \gamma}^{\prime}(0)- \Pi_{\gamma \gamma}^{\prime}(0)\bigg]\,, \\[0.2cm]
  T &= \frac{1}{\alpha}\bigg[\frac{ \Pi_{W W}(0)}{ m_{W}^{2}}-\frac{ \Pi_{Z Z}(0)}{m_{Z}^{2}}\bigg]\,, \\[0.2cm]
  U &= \frac{4 s_{W}^{2}}{\alpha}\bigg[\frac{\Pi_{W W}\left(m_{W}^{2}\right)- \Pi_{W W}(0)}{m_{W}^{2}}
      \nn\\
      & \qquad\qquad-\frac{c_W}{s_W}\Pi_{Z\gamma}^\prime(0)- \Pi_{\gamma \gamma}^\prime(0)\bigg]-S\,, 
\end{align}
where $\Pi_{XY}$ denotes the NP contribution to the vacuum polarization of the gauge bosons with $X,Y=W,Z,\gamma$, $c_W=\cos\theta_W$, and $\alpha$ is the fine structure constant. Then, the $W$-boson mass shift induced by the oblique corrections can be expressed as~\cite{Peskin:1991sw}
\begin{align}
\Delta m_{W}^{2}=\frac{\alpha c_{W}^{2} m_{Z}^{2}}{c_{W}^{2}-s_{W}^{2}}\left[-\frac{S}{2}+c_{W}^{2} T+\frac{c_{W}^{2}-s_{W}^{2}}{4 s_{W}^{2}} U\right].
\end{align}
Thus, to make an explanation of the discrepancy between the latest CDF $m_W$ measurement and the SM expectation, one needs a global EW fit to the oblique parameters $S$, $T$ and $U$, which encode the potential NP effects.

In the model introduced in section~\ref{sec:Zp}, extra contributions to the vacuum polarization of gauge bosons arise from the modified fermion-gauge couplings that are characterized by the mixing angle $\theta_L$, as well as from the loops involving the top partner (cf. eqs.~\eqref{eq:int1}--\eqref{eq:int4}). Their contributions to the oblique parameters can be written as
\begin{align}
  S &= \frac{s_L^2}{12\pi} \Big[ K_1(y_t, y_T) + 3 c_L^2 K_2 (y_t, y_T) \Big], \label{eq:S} \\[0.2cm]
  T &=  \frac{3s_L^2}{16\pi s_W^2} \bigg[x_T-x_t - c_L^2\Big(x_T+x_t +\frac{2x_tx_T}{x_T-x_t}\ln\frac{x_t}{x_T} \Big)\bigg], \label{eq:T} \\[0.2cm]
  U &= \frac{s_L^2}{12\pi} \Big[ K_3(x_t,y_t)-K_3(x_T,y_T)\Big] -S, \label{eq:U}
\end{align}
with $x_q=m_q^2/m_W^2$ and $y_q=m_q^2/m_Z^2$ for $q=t,T$. Explicit expressions of the loop functions $K_{1,2,3}(x,y)$ are listed in the appendix. In the NP model considered, the oblique parameters $S$, $T$ and $U$ are solely determined by the two NP parameters $\theta_L$ and $m_T$.

\section{$b \to s \ell^+ \ell^-$ transitions}
\label{sec:b2s}

\begin{figure}
  \centering
  \includegraphics[width=0.241\linewidth]{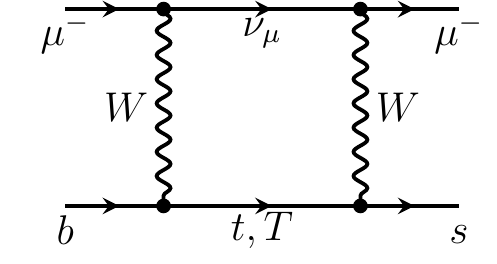}
  \includegraphics[width=0.241\linewidth]{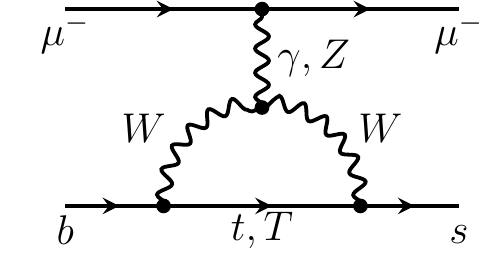}
  \includegraphics[width=0.241\linewidth]{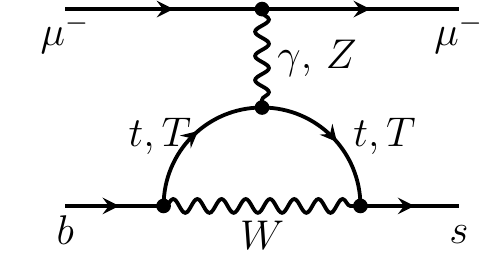}
  \includegraphics[width=0.241\linewidth]{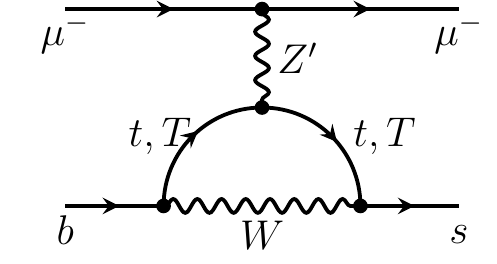}
  \caption{One-loop Feynman diagrams for the $b \to s \mu^+\mu^-$ transitions, corresponding to the $W$-box, as well as the $\gamma$-, $Z$-, and $\Zp$-penguin contributions, respectively.}
  \label{fig:b2s:diagram}
\end{figure}

The FCNC processes mediated by the quark-level $b \to s \ell^+ \ell^-$ transitions, such as the rare decays $B_s \to \ell^+ \ell^-$, $B \to K^{(*)} \ell^+ \ell^- $ and $B_s \to \phi \ell^+ \ell^-$, can also provide promising probes of potential NP effects~\cite{Bifani:2018zmi,Albrecht:2021tul,London:2021lfn}. Specific to the model introduced in section~\ref{sec:Zp}, the NP contributions to the $b \to s \ell^+ \ell^-$ transitions start at the one-loop level, with the relevant Feynman diagrams shown in figure~\ref{fig:b2s:diagram}. Here the NP effects can arise from the modified SM box and penguin diagrams that are featured by the mixing angle $\theta_L$, as well as from the penguin diagrams involving the top partner and/or $\Zp$ boson. 

By an explicit calculation, we find that these NP contributions affect only the Wilson coefficients $\mC_9$ and $\mC_{10}$ in the low-energy effective weak Hamiltonian governing the $b \to s \mu^+ \mu^-$ transitions~\cite{Buchalla:1995vs}
\begin{align}
\mathcal{H}_{\rm eff} \supset -\frac{4 G_{F}}{\sqrt{2}} V_{t b} V_{t s}^{*} \frac{\alpha}{4 \pi}\left(\mC_{9}^{\mu} \mathcal{O}_{9}^{\mu}+ \mC_{10}^{\mu} \mathcal{O}_{10}^{\mu}\right)+\text { h.c.}\,,
\end{align}
with the four-fermion operators $\mathcal{O}_{9}=\left(\bar{s} \gamma^{\mu} P_{L} b\right)\left(\bar\mu \gamma_{\mu} \mu\right)$ and $\mathcal{O}_{10}=\left(\bar{s} \gamma^{\mu} P_{L} b\right)\left(\bar\mu \gamma_{\mu} \gamma_{5} \mu\right)$. Explicitly, the NP contributions to $\mC_9$ and $\mC_{10}$ can be written, respectively, as
\begin{align}
  \mC_9^{\rm NP}&= s_L^2 I_1+ s_L^2 \left(1-\frac{1}{4s_W^2}\right) \left(I_2 + c_L^2 I_3 \right)+ \Delta\mC_+^\Zp\,, \label{eq:WC1}\\[0.15cm]
  \mC_{10}^{\rm NP}&= \frac{s_L^2}{4s_W^2}\left(I_2 + c_L^2 I_3 \right)+ \Delta\mC_-^\Zp\,, \label{eq:WC2}
\end{align}
with the $\Zp$-penguin contributions given by
\begin{align}
  \Delta\mC_{\pm}^\Zp = \frac{(g_L\pm g_R)q_t g_t}{e^2} \frac{m_W^2}{m_\Zp^2}c_L^2s_R^2 \left( I_4 -\frac{c_L^2}{c_R^2} I_5 \right),
\end{align}
where $e=\sqrt{4\pi\alpha}$ is the electromagnetic coupling constant. The loop integrals $I_{1-5}$ are all functions of the masses $m_t$ and $m_T$, whose explicit expressions can be found in the appendix. In the limit of $\theta_L\to 0$ or $\theta_R \to 0$, the NP Wilson coefficients $\mC_9^{\rm NP}$ and $\mC_{10}^{\rm NP}$ will vanish.

Here, the NP contributions to the Wilson coefficients $\mC_9^{\rm NP}$ and $\mC_{10}^{\rm NP}$ are calculated in the unitary gauge. We implement our computation in two setups by making use of different packages including \texttt{FeynRules}~\cite{Alloul:2013bka}, \texttt{FeynArts}~\cite{Hahn:2000kx}, \texttt{FeynCalc}~\cite{Mertig:1990an,Shtabovenko:2016sxi,Shtabovenko:2020gxv}, \texttt{Package-X}~\cite{Patel:2016fam}, as well as some in-house routines. After expanding in terms of the ratio $m_W^2/m_T^2$ and keeping only the leading terms in $s_R=\sin\theta_R$, our results are in agreement with those obtained in ref.~\cite{Fox:2018ldq}.

\section{Numerical analysis}
\label{sec:numerics}

\subsection{$W$-boson mass and global EW fit}

\begin{figure}[t]
  \centering
  \includegraphics[width=0.48\textwidth]{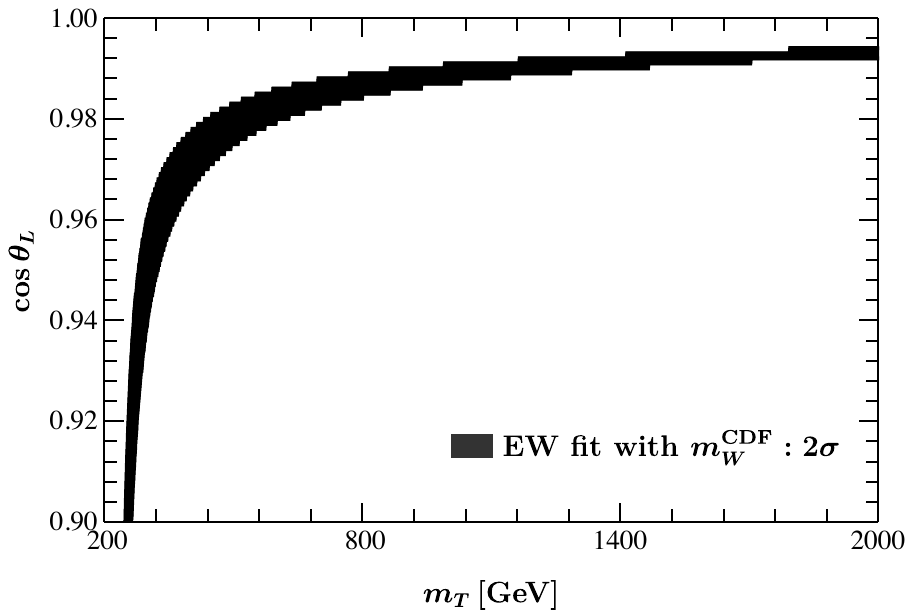}
  \caption{Allowed parameter space in the $(m_T, \cos\theta_L)$ plane at $2\sigma$ level from the global EW fit with the latest CDF measurement of the $W$-boson mass.}
  \label{fig:EWPT}
\end{figure}

As all the three oblique parameters $S$, $T$ and $U$ could be affected by the NP effects (cf. eqs.~\eqref{eq:S}--\eqref{eq:U}), a global EW fit is required to explain the latest CDF $m_W$ measurement~\cite{CDF:2022hxs}. To this end, several groups have recently updated the global EW fit on the $S$, $T$ and $U$ parameters by including the latest CDF measurement of the $W$-bosn mass~\cite{Strumia:2022qkt,Asadi:2022xiy,Gu:2022htv,Lu:2022bgw,deBlas:2022hdk}. Here, we adopt the result obtained in ref.~\cite{Lu:2022bgw},\footnote{It is found that using the global EW fit performed in ref.~\cite{deBlas:2022hdk} with the package \texttt{HEPfit}~\cite{DeBlas:2019ehy} does not make substantial changes to our numerical results.} which is based on the package \texttt{Gfitter}~\cite{Flacher:2008zq,Baak:2014ora,Haller:2018nnx}. The resulting values of the $S$, $T$ and $U$ parameters as well as the correlation matrix read~\cite{Lu:2022bgw}
\begin{align}
  \begin{array}{l}
    S=0.06 \pm 0.10,\\[0.15cm]
    T=0.11 \pm 0.12,\\[0.15cm]
    U=0.14 \pm 0.09,
  \end{array}
  \quad
  {\rm corr.=}
  \begin{bmatrix}
    1.00 & 0.90 & -0.59\\[0.15cm]
    & 1.00 & -0.85 \\[0.15cm]
    & & \hphantom{-}1.00
  \end{bmatrix}.
\end{align}
As described in section~\ref{sec:EWPT}, the NP contributions to the $S$, $T$ and $U$ parameters depend only on the top-partner mass $m_T$ and the mixing angle $\theta_L$. To avoid large modification of the $\bar t b W^+$ coupling, we require $\cos\theta_L>0.9$ in the following. This range is also consistent with that derived from the top-Higgs coupling through global fits to the SM Higgs properties~\cite{Fox:2011qd,Fox:2018ldq}. After considering the constraints from the global EW fit, we find that there still exist parameter regions allowed at $2\sigma$ level, which is shown in figure~\ref{fig:EWPT}. Numerically, $\cos\theta_L=0.98\sim0.99$ is allowed for $500<m_T<2000\GeV$.

Such a small mixing angle suppresses the NP contributions from the $W$-box, $\gamma$- and $Z$-penguin diagrams, because they are all proportional to $\sin^2\theta_L$, as can be seen from eqs.~\eqref{eq:WC1} and \eqref{eq:WC2}. However, the $\Zp$-penguin diagrams do not suffer from this suppression and may affect the $b \to s \ell^+ \ell^-$ processes, as will be exploited in the next subsection.

\subsection{$b \to s \ell^+ \ell^-$ anomalies}

In the model considered, the NP affects only the Wilson coefficients $\mC_9$ and $\mC_{10}$ and, in order to find the parameter regions required to explain the $b \to s \ell^+ \ell^-$ anomalies, we should perform a global fit to the $b \to s \ell^+ \ell^-$ observables, including, but not limited to, the lepton-flavour-universality-violating ratios $R_{K^{(*)}}$~\cite{LHCb:2017avl,LHCb:2021trn,LHCb:2021lvy} and the angular observables in $B \to K^* \mu^+ \mu^-$ decays~\cite{LHCb:2020lmf,LHCb:2020gog}. When performing the global fit, we follow the same prescription as in ref.~\cite{Altmannshofer:2014rta} and use the package \texttt{flavio}~\cite{Straub:2018kue}. More details can be found in ref.~\cite{Li:2021qyo}. We refer to refs.~\cite{Geng:2021nhg,Altmannshofer:2021qrr,Cornella:2021sby,Alguero:2021anc,Hurth:2021nsi} for the latest global fit by other groups. 

The NP contributions to the $b \to s \ell^+ \ell^-$ transitions depend on the mixing angles $\theta_{L,R}$, the top-partner mass $m_T$, the $U(1)^\prime$ charge $q_t$, the $\Zp$ mass $m_\Zp$, the $\bar t t\Zp$ coupling $g_t$, as well as the $\mu^+ \mu^-\Zp$ couplings $g_\mu^{L,R}$. From eqs.~\eqref{eq:WC1} and \eqref{eq:WC2}, one can see that the $\Zp$-related parameters always appear as the product $q_tg_tg_\mu^{L,R}/\mZp^2$. Thus, without loss of generality, we take in the numerical analysis $q_t=1$, $g_t=1$ and $m_\Zp=200\GeV$, while keeping the effective $\mu^+ \mu^-\Zp$ couplings $g_\mu^{L,R}$ as free parameters. After considering the relation in eq.~\eqref{eq:theta_L}, the independent NP parameters can be chosen as $(\cos\theta_L, m_T, g_\mu^L, g_\mu^R)$.

\begin{figure*}[t]
	\centering
	\includegraphics[width=0.245\textwidth]{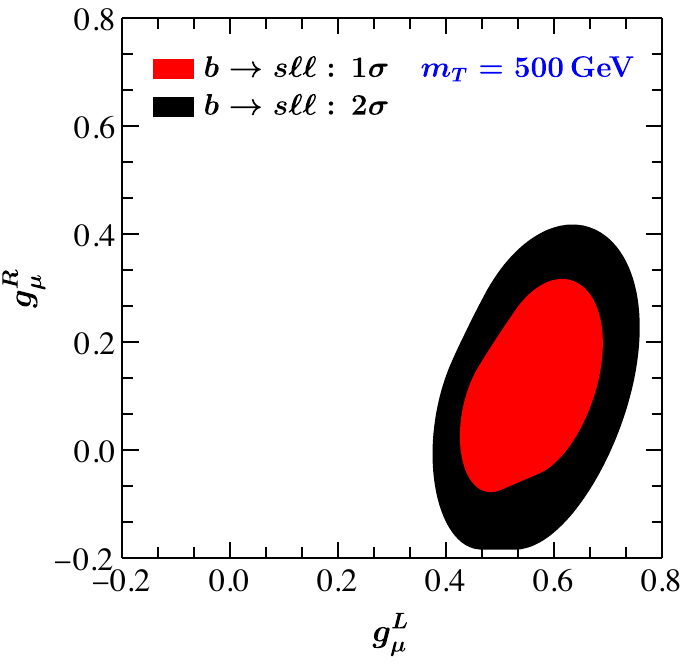}
	\includegraphics[width=0.245\textwidth]{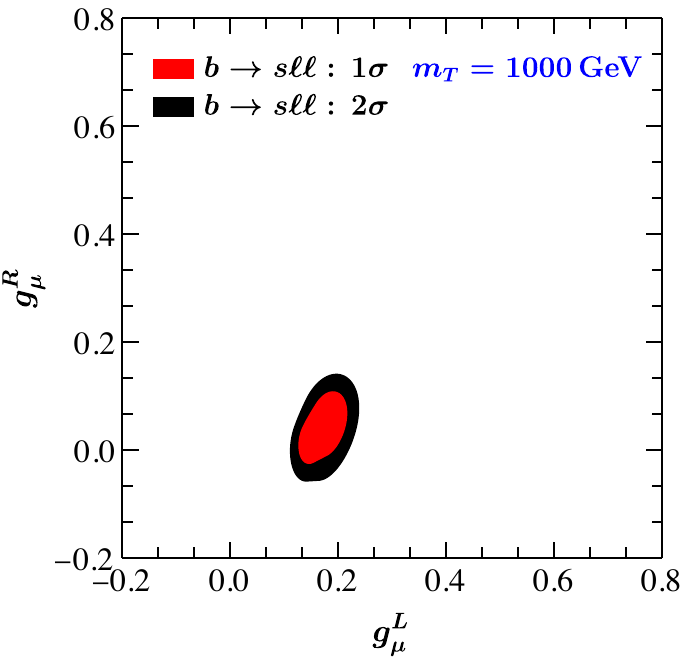}
	\includegraphics[width=0.245\textwidth]{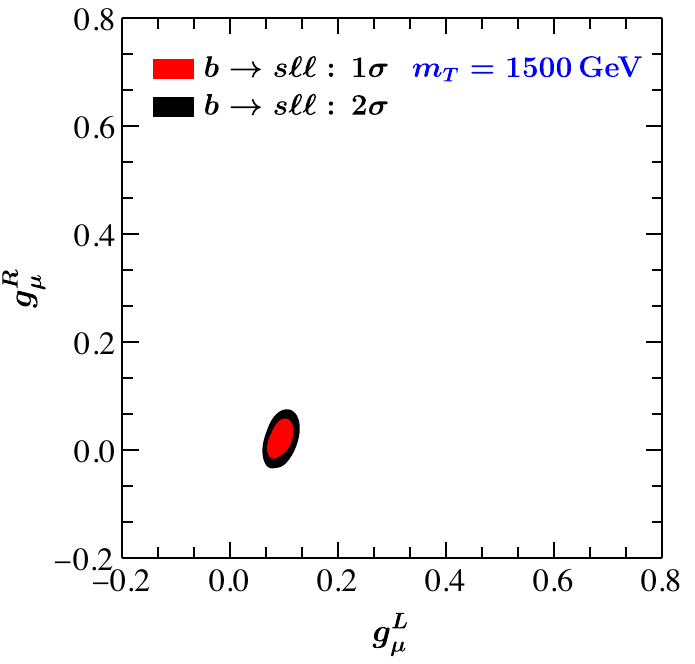}
	\includegraphics[width=0.245\textwidth]{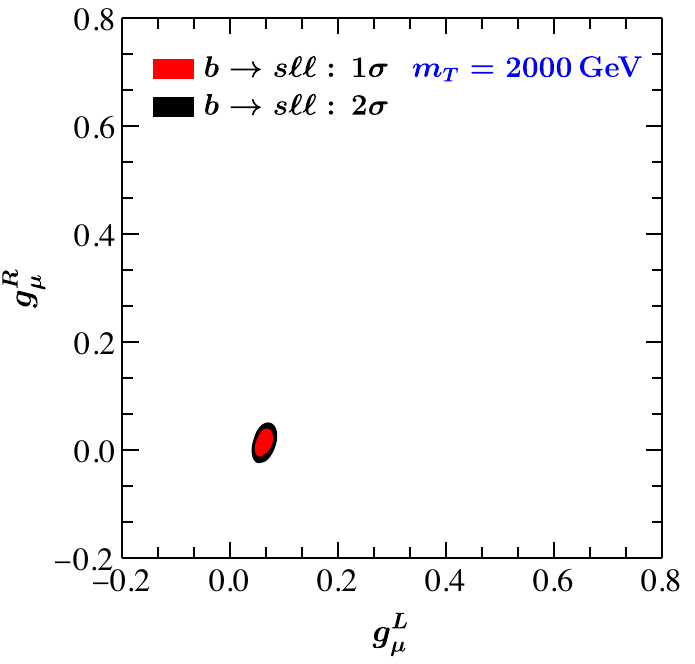}
	\caption{Constraints on the effective $\mu^+ \mu^-\Zp$ couplings $(g_\mu^L,g_\mu^R)$ by the $b \to s \ell^+ \ell^-$ processes, in the $2\sigma$ allowed regions of the $(\cos\theta_L, m_T)$ plane obtained from the global EW fit with the latest CDF $m_W$ measurement. The allowed regions are shown in red ($1\sigma$) and black ($2\sigma$) for $m_T=500$, $1000$, $1500$, $2000\GeV$, respectively. Here, without loss of generality, $q_t=1$, $g_t=1$ and $m_\Zp=200\GeV$ have been taken.}
	\label{fig:b2s}
\end{figure*}

With the $2\sigma$ allowed regions of $\cos\theta_L$ and $m_T$ obtained from the global EW fit (cf. figure~\ref{fig:EWPT}), constraints on the NP parameters $g_\mu^{L,R}$ can then be derived from the $b \to s \ell^+ \ell^-$ processes. In figure~\ref{fig:b2s}, we show the allowed regions of $(g_\mu^L, g_\mu^R)$ for $m_T=500$, $1000$, $1500$, $2000\GeV$, respectively. It can be seen that, even after considering the latest CDF measurement of the $W$-boson mass, the model introduced in section~\ref{sec:Zp} can still accommodate the $b \to s \ell^+ \ell^-$ anomalies. For a $\Zp$ boson with mass of hundreds of $\GeV$, its couplings to the top and the muon are both less than $\mO(1)$, being therefore safely in the perturbative region. Furthermore, smaller $\Zp$ couplings to the top and the muon are required in the case of a heavier top partner.

\subsection{Collider searches}
\label{sec:otherprocesses}

Direct searches for single and pair productions of the vector-like top partner have been performed at the LHC, and strong bounds on the mass and mixing angle of the top partner have been derived~\cite{ATLAS:2018ziw,CMS:2022yxp,Aguilar-Saavedra:2013qpa,delaTorreTrishaFarooque:2022vqc}. However, most of these searches assume that the top partner decays exclusively into the SM particles, \textit{i.e.}, $T \to bW/tZ/th$. In our case, however, $T \to t \Zp$ is the dominant decay channel in the range $\mZp<m_T-m_t$, which could relax the constraints from these direct searches~\cite{Serra:2015xfa,Anandakrishnan:2015yfa,Bizot:2018tds}.

In the model considered, the $\Zp$ boson couples exclusively to the top quark and the top partner. Collider phenomenologies of such a top-philic $\Zp$ have been investigated in refs.~\cite{Greiner:2014qna,Cox:2015afa,Kim:2016plm,Kamenik:2017tnu,Fox:2018ldq}. At the LHC, the dominant production channels include $pp \to t \bar t \Zp$ at the tree and $pp \to j \Zp$ at the one-loop level~\cite{Fox:2018ldq}. The $\Zp \to \mu^+ \mu^-$ decay makes the dimuon resonance searches~\cite{ATLAS:2019erb,CMS:2021ctt} sensitive to this scenario. In addition, searches for multi-top final states could also provide important probe of the $\Zp$ boson as well as the top partner~\cite{Fox:2018ldq}. A complete analysis of the collider searches may provide interesting constraints, which is however beyond the scope of this work.

\section{Conclusion}
\label{sec:conclusion}

The latest CDF measurement of the $W$-boson mass shows about $7\sigma$ deviation from the SM expectation obtained by a global EW fit. We have pointed out that the latest CDF $m_W$ measurement and the $b \to s \ell^+ \ell^-$ anomalies could be simultaneously accommodated by introducing additional fermions, which have the same quantum numbers as of the SM quarks. As a simple example, the model introduced in ref.~\cite{Kamenik:2017tnu}, which is characterized by a vector-like top partner with an additional $U(1)^\prime$ gauge symmetry, is considered. We have calculated its contributions to the oblique parameters $S$, $T$ and $U$ in the global EW fit as well as the Wilson coefficients $\mC_9$ and $\mC_{10}$ in the $b \to s \ell^+ \ell^-$ transitions. Our numerical results show that the top-partner loops in the vacuum polarization of gauge bosons can explain the latest CDF $m_W$ measurement and, at the same time, the $\Zp$-penguin diagrams involving the top partner can account for the $b \to s \ell^+ \ell^-$ anomalies. Both the $\Zp$ boson and the top partner can be as light as of a few hundreds of $\GeV$ and thus may be accessible at the LHC Run III and its upgrade. Therefore, our work has demonstrated that the latest CDF $W$-boson mass shift could be related to the $b \to s \ell^+ \ell^-$ anomalies, which is expected to inspire more complete model buildings for unified theories containing new fermions.

\section*{Note Added}
As this paper was being finalized, refs.~\cite{Lee:2022nqz,Crivellin:2022fdf,Endo:2022kiw,Balkin:2022glu,Cao:2022mif,Baek:2022agi,Borah:2022zim} appeared that also consider possible explanations of the latest CDF $m_W$ measurement with vector-like top partners.

\section*{Acknowledgements}
This work is supported by the National Natural Science Foundation of China under Grant Nos. 12135006, 12075097, and 11805077, as well as by the self-determined research funds of CCNU from the colleges’ basic research and operation of MOE under Grant Nos. CCNU19TD012 and CCNU20TS007. XY is also supported in part by the Startup Research Funding from CCNU.

\begin{appendices}
	
\renewcommand{\theequation}{A.\arabic{equation}}	

\section*{Appendix. Loop functions}

The loop functions $K_{1,2,3}(x,y)$ in the oblique parameters $S$, $T$ and $U$ in eqs.~\eqref{eq:S}--\eqref{eq:U} are listed below
\begin{align}
  K_1(x,y) &=  -44x + 2\ln x + f_{-9}(x,x) - (x \to y),
  \\[0.2cm]
   K_2(x,y) &= (x-y)^2 -\left[(x-y)^3 - 3(x-y) -\frac{6xy}{x-y} \right]\ln x \nn \\
   & -3x - f_3(x,x) + f_3(x,y) + (x \leftrightarrow y), \\[0.2cm]
  K_3(x,y) &= -3 x - 6 x^2 -32 y +(8-18x+6x^3)\ln x  \nn \\
  & - (12 - 18 x+ 6x^3) \ln(x-1) + 4 f_0(y, y),
\end{align}
with the function
\begin{align}
 f_n(x,y) &= -\left[(x+y-1)^2-4xy-3+n(x+y)\right] \nn \\
  & \hspace{-0.5cm} \times \sqrt{4xy-(x+y-1)^2} \cos^{-1}\left(\frac{x+y-1}{\sqrt{4xy}}\right).
\end{align}

Explicit expressions of the loop functions in the Wilson coefficients $\mC_9^{\rm NP}$ and $\mC_{10}^{\rm NP}$ in eqs.~\eqref{eq:WC1} and \eqref{eq:WC2} are given, respectively, by
\begin{align}
    I_1 & = f_t\left(-\frac{17}{9},-\frac{7}{36},\frac{1}{6}\right) \nn \\
      & +\left[-\frac{1}{6}+f_t\left(2,2,\frac{1}{9},-\frac{1}{6}\right)\right]\ln x_t - (t \to T), \\[0.2cm]
  I_2 & = \frac{x_t}{2}-f_t\left(\frac{3}{2}\right)+ \left[\frac{3}{2}+f_t\left(3,\frac{3}{2}\right) \right]\ln x_t - (t \to T), \\[0.2cm]
  I_3 & = \frac{x_t+x_T}{2}+\frac{x_tx_T}{x_T-x_t}\ln\frac{x_t}{x_T}, \\[0.15cm]
  I_4 & = \frac{3x_t}{4} \left[ \frac{1}{3}-f_t(1) +2g(x_T,x_t)\ln x_t + (t \leftrightarrow T) \right],
  \\[0.20cm]
  I_5 & = \frac{x_t^2}{2}\left[\frac{1}{x_t} +\frac{2\ln x_t}{x_T-x_t} + (t \leftrightarrow T) \right],
\end{align}
with $x_t=m_t^2/m_W^2$, $x_T=m_T^2/m_W^2$, and the functions
\begin{align}
  g(x,y)=\frac{y(4-8y+y^2)+x(4-2y+y^2)}{6(x-y)(y-1)^2},  \\[0.2cm]
  f_q(a_1, a_2, a_3, \dotsc, a_n) = \sum_{i=1}^n\frac{a_i}{(x_q-1)^i}. 
\end{align}

\end{appendices}

\bibliographystyle{elsarticle-num}

\bibliography{ref}

\begin{thebibliography}{10}
\expandafter\ifx\csname url\endcsname\relax
  \def\url#1{\texttt{#1}}\fi
\expandafter\ifx\csname urlprefix\endcsname\relax\def\urlprefix{URL }\fi
\expandafter\ifx\csname href\endcsname\relax
  \def\href#1#2{#2} \def\path#1{#1}\fi

\bibitem{Zyla:2020zbs}
P.~Zyla, et~al., {Review of Particle Physics}, PTEP 2020~(8) (2020) 083C01, and
  2021 update.
\newblock \href {https://doi.org/10.1093/ptep/ptaa104}
  {\path{doi:10.1093/ptep/ptaa104}}.

\bibitem{ATLAS:2012yve}
G.~Aad, et~al., {Observation of a new particle in the search for the Standard
  Model Higgs boson with the ATLAS detector at the LHC}, Phys. Lett. B 716
  (2012) 1--29.
\newblock \href {http://arxiv.org/abs/1207.7214} {\path{arXiv:1207.7214}},
  \href {https://doi.org/10.1016/j.physletb.2012.08.020}
  {\path{doi:10.1016/j.physletb.2012.08.020}}.

\bibitem{CMS:2012qbp}
S.~Chatrchyan, et~al., {Observation of a New Boson at a Mass of 125 GeV with
  the CMS Experiment at the LHC}, Phys. Lett. B 716 (2012) 30--61.
\newblock \href {http://arxiv.org/abs/1207.7235} {\path{arXiv:1207.7235}},
  \href {https://doi.org/10.1016/j.physletb.2012.08.021}
  {\path{doi:10.1016/j.physletb.2012.08.021}}.

\bibitem{ZurbanoFernandez:2020cco}
I.~Zurbano~Fernandez, et~al., {High-Luminosity Large Hadron Collider (HL-LHC):
  Technical design report} 10/2020 (12 2020).
\newblock \href {https://doi.org/10.23731/CYRM-2020-0010}
  {\path{doi:10.23731/CYRM-2020-0010}}.

\bibitem{Bifani:2018zmi}
S.~Bifani, S.~Descotes-Genon, A.~Romero~Vidal, M.-H. Schune, {Review of Lepton
  Universality tests in $B$ decays}, J. Phys. G 46~(2) (2019) 023001.
\newblock \href {http://arxiv.org/abs/1809.06229} {\path{arXiv:1809.06229}},
  \href {https://doi.org/10.1088/1361-6471/aaf5de}
  {\path{doi:10.1088/1361-6471/aaf5de}}.

\bibitem{Albrecht:2021tul}
J.~Albrecht, D.~van Dyk, C.~Langenbruch, {Flavour anomalies in heavy quark
  decays}, Prog. Part. Nucl. Phys. 120 (2021) 103885.
\newblock \href {http://arxiv.org/abs/2107.04822} {\path{arXiv:2107.04822}},
  \href {https://doi.org/10.1016/j.ppnp.2021.103885}
  {\path{doi:10.1016/j.ppnp.2021.103885}}.

\bibitem{London:2021lfn}
D.~London, J.~Matias, {$B$ Flavour Anomalies: 2021 Theoretical Status Report}
  (10 2021).
\newblock \href {http://arxiv.org/abs/2110.13270} {\path{arXiv:2110.13270}},
  \href {https://doi.org/10.1146/annurev-nucl-102020-090209}
  {\path{doi:10.1146/annurev-nucl-102020-090209}}.

\bibitem{Hollik:1988ii}
W.~F.~L. Hollik, {Radiative Corrections in the Standard Model and their Role
  for Precision Tests of the Electroweak Theory}, Fortsch. Phys. 38 (1990)
  165--260.
\newblock \href {https://doi.org/10.1002/prop.2190380302}
  {\path{doi:10.1002/prop.2190380302}}.

\bibitem{Langacker:1991zr}
P.~Langacker, M.-x. Luo, A.~K. Mann, {High precision electroweak experiments: A
  Global search for new physics beyond the standard model}, Rev. Mod. Phys. 64
  (1992) 87--192.
\newblock \href {https://doi.org/10.1103/RevModPhys.64.87}
  {\path{doi:10.1103/RevModPhys.64.87}}.

\bibitem{Erler:2019hds}
J.~Erler, M.~Schott, {Electroweak Precision Tests of the Standard Model after
  the Discovery of the Higgs Boson}, Prog. Part. Nucl. Phys. 106 (2019)
  68--119.
\newblock \href {http://arxiv.org/abs/1902.05142} {\path{arXiv:1902.05142}},
  \href {https://doi.org/10.1016/j.ppnp.2019.02.007}
  {\path{doi:10.1016/j.ppnp.2019.02.007}}.

\bibitem{CDF:2022hxs}
T.~Aaltonen, et~al., {High-precision measurement of the W boson mass with the
  CDF II detector}, Science 376~(6589) (2022) 170--176.
\newblock \href {https://doi.org/10.1126/science.abk1781}
  {\path{doi:10.1126/science.abk1781}}.

\bibitem{LHCb:2021bjt}
R.~Aaij, et~al., {Measurement of the W boson mass}, JHEP 01 (2022) 036.
\newblock \href {http://arxiv.org/abs/2109.01113} {\path{arXiv:2109.01113}},
  \href {https://doi.org/10.1007/JHEP01(2022)036}
  {\path{doi:10.1007/JHEP01(2022)036}}.

\bibitem{Strumia:2022qkt}
A.~Strumia, {Interpreting electroweak precision data including the $W$-mass CDF
  anomaly} (4 2022).
\newblock \href {http://arxiv.org/abs/2204.04191} {\path{arXiv:2204.04191}}.

\bibitem{Asadi:2022xiy}
P.~Asadi, C.~Cesarotti, K.~Fraser, S.~Homiller, A.~Parikh, {Oblique Lessons
  from the $W$ Mass Measurement at CDF II} (4 2022).
\newblock \href {http://arxiv.org/abs/2204.05283} {\path{arXiv:2204.05283}}.

\bibitem{Gu:2022htv}
J.~Gu, Z.~Liu, T.~Ma, J.~Shu, {Speculations on the W-Mass Measurement at CDF}
  (4 2022).
\newblock \href {http://arxiv.org/abs/2204.05296} {\path{arXiv:2204.05296}}.

\bibitem{Lu:2022bgw}
C.-T. Lu, L.~Wu, Y.~Wu, B.~Zhu, {Electroweak Precision Fit and New Physics in
  light of $W$ Boson Mass} (4 2022).
\newblock \href {http://arxiv.org/abs/2204.03796} {\path{arXiv:2204.03796}}.

\bibitem{deBlas:2022hdk}
J.~de~Blas, M.~Pierini, L.~Reina, L.~Silvestrini, {Impact of the recent
  measurements of the top-quark and W-boson masses on electroweak precision
  fits} (4 2022).
\newblock \href {http://arxiv.org/abs/2204.04204} {\path{arXiv:2204.04204}}.

\bibitem{Fan:2022yly}
J.~Fan, L.~Li, T.~Liu, K.-F. Lyu, {$W$-Boson Mass, Electroweak Precision Tests
  and SMEFT} (4 2022).
\newblock \href {http://arxiv.org/abs/2204.04805} {\path{arXiv:2204.04805}}.

\bibitem{Liu:2022jdq}
X.~Liu, S.-Y. Guo, B.~Zhu, Y.~Li, {Unifying gravitational waves with $W$ boson,
  FIMP dark matter, and Majorana Seesaw mechanism} (4 2022).
\newblock \href {http://arxiv.org/abs/2204.04834} {\path{arXiv:2204.04834}}.

\bibitem{Bagnaschi:2022whn}
E.~Bagnaschi, J.~Ellis, M.~Madigan, K.~Mimasu, V.~Sanz, T.~You, {SMEFT Analysis
  of $m_{W}$} (4 2022).
\newblock \href {http://arxiv.org/abs/2204.05260} {\path{arXiv:2204.05260}}.

\bibitem{Paul:2022dds}
A.~Paul, M.~Valli, {Violation of custodial symmetry from W-boson mass
  measurements} (4 2022).
\newblock \href {http://arxiv.org/abs/2204.05267} {\path{arXiv:2204.05267}}.

\bibitem{Heo:2022dey}
Y.~Heo, D.-W. Jung, J.~S. Lee, {Impact of the CDF $W$-mass anomaly on two Higgs
  doublet model} (4 2022).
\newblock \href {http://arxiv.org/abs/2204.05728} {\path{arXiv:2204.05728}}.

\bibitem{Zheng:2022irz}
M.-D. Zheng, F.-Z. Chen, H.-H. Zhang, {The $W\ell\nu$-vertex corrections to
  W-boson mass in the R-parity violating MSSM} (4 2022).
\newblock \href {http://arxiv.org/abs/2204.06541} {\path{arXiv:2204.06541}}.

\bibitem{Almeida:2022lcs}
E.~d.~S. Almeida, A.~Alves, O.~J.~P. Eboli, M.~C. Gonzalez-Garcia, {Impact of
  CDF-II measurement of $M_W$ on the electroweak legacy of the LHC Run II} (4
  2022).
\newblock \href {http://arxiv.org/abs/2204.10130} {\path{arXiv:2204.10130}}.

\bibitem{Cheng:2022aau}
Y.~Cheng, X.-G. He, F.~Huang, J.~Sun, Z.-P. Xing, {Dark photon kinetic mixing
  effects for CDF W mass excess} (4 2022).
\newblock \href {http://arxiv.org/abs/2204.10156} {\path{arXiv:2204.10156}}.

\bibitem{Chen:2022ocr}
T.-K. Chen, C.-W. Chiang, K.~Yagyu, {Explanation of the $W$ mass shift at CDF
  II in the Georgi-Machacek Model} (4 2022).
\newblock \href {http://arxiv.org/abs/2204.12898} {\path{arXiv:2204.12898}}.

\bibitem{Alguero:2022est}
M.~Alguer\'o, A.~Crivellin, C.~A. Manzari, J.~Matias, {Importance of
  $Z-Z^\prime$ Mixing in $b\to s\ell^+\ell^-$ and the $W$ mass} (1 2022).
\newblock \href {http://arxiv.org/abs/2201.08170} {\path{arXiv:2201.08170}}.

\bibitem{Zhou:2022cql}
Q.~Zhou, X.-F. Han, {The CDF W-mass, muon g-2, and dark matter in a
  $U(1)_{L_\mu-L_\tau}$ model with vector-like leptons} (4 2022).
\newblock \href {http://arxiv.org/abs/2204.13027} {\path{arXiv:2204.13027}}.

\bibitem{Popov:2022ldh}
O.~Popov, R.~Srivastava, {The Triplet Dirac Seesaw in the View of the Recent
  CDF-II W Mass Anomaly} (4 2022).
\newblock \href {http://arxiv.org/abs/2204.08568} {\path{arXiv:2204.08568}}.

\bibitem{Ghorbani:2022vtv}
K.~Ghorbani, P.~Ghorbani, {$W$-Boson Mass Anomaly from Scale Invariant 2HDM} (4
  2022).
\newblock \href {http://arxiv.org/abs/2204.09001} {\path{arXiv:2204.09001}}.

\bibitem{Bhaskar:2022vgk}
A.~Bhaskar, A.~A. Madathil, T.~Mandal, S.~Mitra, {Combined explanation of
  $W$-mass, muon $g-2$, $R_{K^{(*)}}$ and $R_{D^{(*)}}$ anomalies in a
  singlet-triplet scalar leptoquark model} (4 2022).
\newblock \href {http://arxiv.org/abs/2204.09031} {\path{arXiv:2204.09031}}.

\bibitem{Heckman:2022the}
J.~J. Heckman, {Extra $W$-Boson Mass from a D3-Brane} (4 2022).
\newblock \href {http://arxiv.org/abs/2204.05302} {\path{arXiv:2204.05302}}.

\bibitem{DiLuzio:2022xns}
L.~Di~Luzio, R.~Gr\"ober, P.~Paradisi, {Higgs physics confronts the $M_W$
  anomaly} (4 2022).
\newblock \href {http://arxiv.org/abs/2204.05284} {\path{arXiv:2204.05284}}.

\bibitem{Fan:2022dck}
Y.-Z. Fan, T.-P. Tang, Y.-L.~S. Tsai, L.~Wu, {Inert Higgs Dark Matter for New
  CDF W-boson Mass and Detection Prospects} (4 2022).
\newblock \href {http://arxiv.org/abs/2204.03693} {\path{arXiv:2204.03693}}.

\bibitem{Tang:2022pxh}
T.-P. Tang, M.~Abdughani, L.~Feng, Y.-L.~S. Tsai, Y.-Z. Fan, {NMSSM neutralino
  dark matter for $W$-boson mass and muon $g-2$ and the promising prospect of
  direct detection} (4 2022).
\newblock \href {http://arxiv.org/abs/2204.04356} {\path{arXiv:2204.04356}}.

\bibitem{Borah:2022obi}
D.~Borah, S.~Mahapatra, D.~Nanda, N.~Sahu, {Type II Dirac Seesaw with
  Observable $\Delta N_{eff}$ in the light of W-mass Anomaly} (4 2022).
\newblock \href {http://arxiv.org/abs/2204.08266} {\path{arXiv:2204.08266}}.

\bibitem{Sakurai:2022hwh}
K.~Sakurai, F.~Takahashi, W.~Yin, {Singlet extensions and W boson mass in the
  light of the CDF II result} (4 2022).
\newblock \href {http://arxiv.org/abs/2204.04770} {\path{arXiv:2204.04770}}.

\bibitem{Athron:2022qpo}
P.~Athron, A.~Fowlie, C.-T. Lu, L.~Wu, Y.~Wu, B.~Zhu, {The $W$ boson Mass and
  Muon $g-2$: Hadronic Uncertainties or New Physics?} (4 2022).
\newblock \href {http://arxiv.org/abs/2204.03996} {\path{arXiv:2204.03996}}.

\bibitem{Athron:2022isz}
P.~Athron, M.~Bach, D.~H.~J. Jacob, W.~Kotlarski, D.~St\"ockinger, A.~Voigt,
  {Precise calculation of the W boson pole mass beyond the Standard Model with
  FlexibleSUSY} (4 2022).
\newblock \href {http://arxiv.org/abs/2204.05285} {\path{arXiv:2204.05285}}.

\bibitem{Lee:2022nqz}
H.~M. Lee, K.~Yamashita, {A Model of Vector-like Leptons for the Muon $g-2$ and
  the $W$ Boson Mass} (4 2022).
\newblock \href {http://arxiv.org/abs/2204.05024} {\path{arXiv:2204.05024}}.

\bibitem{Crivellin:2022fdf}
A.~Crivellin, M.~Kirk, T.~Kitahara, F.~Mescia, {Correlating $t\to cZ$ to the
  $W$ Mass and $B$ Physics with Vector-Like Quarks} (4 2022).
\newblock \href {http://arxiv.org/abs/2204.05962} {\path{arXiv:2204.05962}}.

\bibitem{Endo:2022kiw}
M.~Endo, S.~Mishima, {New physics interpretation of $W$-boson mass anomaly} (4
  2022).
\newblock \href {http://arxiv.org/abs/2204.05965} {\path{arXiv:2204.05965}}.

\bibitem{Balkin:2022glu}
R.~Balkin, E.~Madge, T.~Menzo, G.~Perez, Y.~Soreq, J.~Zupan, {On the
  implications of positive W mass shift} (4 2022).
\newblock \href {http://arxiv.org/abs/2204.05992} {\path{arXiv:2204.05992}}.

\bibitem{Cao:2022mif}
J.~Cao, L.~Meng, L.~Shang, S.~Wang, B.~Yang, {Interpreting the $W$ mass anomaly
  in the vectorlike quark models} (4 2022).
\newblock \href {http://arxiv.org/abs/2204.09477} {\path{arXiv:2204.09477}}.

\bibitem{Baek:2022agi}
S.~Baek, {Implications of CDF $W$-mass and $(g-2)_\mu$ on $U(1)_{L_\mu-L_\tau}$
  model} (4 2022).
\newblock \href {http://arxiv.org/abs/2204.09585} {\path{arXiv:2204.09585}}.

\bibitem{LHCb:2017avl}
R.~Aaij, et~al., {Test of lepton universality with $B^{0} \rightarrow
  K^{*0}\ell^{+}\ell^{-}$ decays}, JHEP 08 (2017) 055.
\newblock \href {http://arxiv.org/abs/1705.05802} {\path{arXiv:1705.05802}},
  \href {https://doi.org/10.1007/JHEP08(2017)055}
  {\path{doi:10.1007/JHEP08(2017)055}}.

\bibitem{LHCb:2021trn}
R.~Aaij, et~al., {Test of lepton universality in beauty-quark decays}, Nature
  Phys. 18~(3) (2022) 277--282.
\newblock \href {http://arxiv.org/abs/2103.11769} {\path{arXiv:2103.11769}},
  \href {https://doi.org/10.1038/s41567-021-01478-8}
  {\path{doi:10.1038/s41567-021-01478-8}}.

\bibitem{LHCb:2021lvy}
R.~Aaij, et~al., {Tests of lepton universality using $B^0\to K^0_S \ell^+
  \ell^-$ and $B^+\to K^{*+} \ell^+ \ell^-$ decays} (10 2021).
\newblock \href {http://arxiv.org/abs/2110.09501} {\path{arXiv:2110.09501}}.

\bibitem{LHCb:2020lmf}
R.~Aaij, et~al., {Measurement of $CP$-Averaged Observables in the
  $B^{0}\rightarrow K^{*0}\mu^{+}\mu^{-}$ Decay}, Phys. Rev. Lett. 125~(1)
  (2020) 011802.
\newblock \href {http://arxiv.org/abs/2003.04831} {\path{arXiv:2003.04831}},
  \href {https://doi.org/10.1103/PhysRevLett.125.011802}
  {\path{doi:10.1103/PhysRevLett.125.011802}}.

\bibitem{LHCb:2020gog}
R.~Aaij, et~al., {Angular Analysis of the $B^{+}\rightarrow
  K^{\ast+}\mu^{+}\mu^{-}$ Decay}, Phys. Rev. Lett. 126~(16) (2021) 161802.
\newblock \href {http://arxiv.org/abs/2012.13241} {\path{arXiv:2012.13241}},
  \href {https://doi.org/10.1103/PhysRevLett.126.161802}
  {\path{doi:10.1103/PhysRevLett.126.161802}}.

\bibitem{Kamenik:2017tnu}
J.~F. Kamenik, Y.~Soreq, J.~Zupan, {Lepton flavor universality violation
  without new sources of quark flavor violation}, Phys. Rev. D 97~(3) (2018)
  035002.
\newblock \href {http://arxiv.org/abs/1704.06005} {\path{arXiv:1704.06005}},
  \href {https://doi.org/10.1103/PhysRevD.97.035002}
  {\path{doi:10.1103/PhysRevD.97.035002}}.

\bibitem{Fox:2011qd}
P.~J. Fox, J.~Liu, D.~Tucker-Smith, N.~Weiner, {An Effective Z'}, Phys. Rev. D
  84 (2011) 115006.
\newblock \href {http://arxiv.org/abs/1104.4127} {\path{arXiv:1104.4127}},
  \href {https://doi.org/10.1103/PhysRevD.84.115006}
  {\path{doi:10.1103/PhysRevD.84.115006}}.

\bibitem{Berger:2012ec}
J.~Berger, J.~Hubisz, M.~Perelstein, {A Fermionic Top Partner: Naturalness and
  the LHC}, JHEP 07 (2012) 016.
\newblock \href {http://arxiv.org/abs/1205.0013} {\path{arXiv:1205.0013}},
  \href {https://doi.org/10.1007/JHEP07(2012)016}
  {\path{doi:10.1007/JHEP07(2012)016}}.

\bibitem{Greiner:2014qna}
N.~Greiner, K.~Kong, J.-C. Park, S.~C. Park, J.-C. Winter, {Model-Independent
  Production of a Top-Philic Resonance at the LHC}, JHEP 04 (2015) 029.
\newblock \href {http://arxiv.org/abs/1410.6099} {\path{arXiv:1410.6099}},
  \href {https://doi.org/10.1007/JHEP04(2015)029}
  {\path{doi:10.1007/JHEP04(2015)029}}.

\bibitem{Cox:2015afa}
P.~Cox, A.~D. Medina, T.~S. Ray, A.~Spray, {Novel collider and dark matter
  phenomenology of a top-philic $Z^\prime$}, JHEP 06 (2016) 110.
\newblock \href {http://arxiv.org/abs/1512.00471} {\path{arXiv:1512.00471}},
  \href {https://doi.org/10.1007/JHEP06(2016)110}
  {\path{doi:10.1007/JHEP06(2016)110}}.

\bibitem{Kim:2016plm}
J.~H. Kim, K.~Kong, S.~J. Lee, G.~Mohlabeng, {Probing TeV scale Top-Philic
  Resonances with Boosted Top-Tagging at the High Luminosity LHC}, Phys. Rev. D
  94~(3) (2016) 035023.
\newblock \href {http://arxiv.org/abs/1604.07421} {\path{arXiv:1604.07421}},
  \href {https://doi.org/10.1103/PhysRevD.94.035023}
  {\path{doi:10.1103/PhysRevD.94.035023}}.

\bibitem{Fox:2018ldq}
P.~J. Fox, I.~Low, Y.~Zhang, {Top-philic $Z'$ forces at the LHC}, JHEP 03
  (2018) 074.
\newblock \href {http://arxiv.org/abs/1801.03505} {\path{arXiv:1801.03505}},
  \href {https://doi.org/10.1007/JHEP03(2018)074}
  {\path{doi:10.1007/JHEP03(2018)074}}.

\bibitem{Crivellin:2020oup}
A.~Crivellin, C.~A. Manzari, M.~Alguero, J.~Matias, {Combined Explanation of
  the $Z\to b\bar{b}$ Forward-Backward Asymmetry, the Cabibbo Angle Anomaly,
  $\tau \to \nu \nu$ and $b \to s \ell^+ \ell^-$ Data}, Phys. Rev. Lett.
  127~(1) (2021) 011801.
\newblock \href {http://arxiv.org/abs/2010.14504} {\path{arXiv:2010.14504}},
  \href {https://doi.org/10.1103/PhysRevLett.127.011801}
  {\path{doi:10.1103/PhysRevLett.127.011801}}.

\bibitem{Crivellin:2015era}
A.~Crivellin, L.~Hofer, J.~Matias, U.~Nierste, S.~Pokorski, J.~Rosiek,
  {Lepton-flavour violating $B$ decays in generic $Z'$ models}, Phys. Rev. D
  92~(5) (2015) 054013.
\newblock \href {http://arxiv.org/abs/1504.07928} {\path{arXiv:1504.07928}},
  \href {https://doi.org/10.1103/PhysRevD.92.054013}
  {\path{doi:10.1103/PhysRevD.92.054013}}.

\bibitem{AristizabalSierra:2015vqb}
D.~Aristizabal~Sierra, F.~Staub, A.~Vicente, {Shedding light on the $b\to s$
  anomalies with a dark sector}, Phys. Rev. D 92~(1) (2015) 015001.
\newblock \href {http://arxiv.org/abs/1503.06077} {\path{arXiv:1503.06077}},
  \href {https://doi.org/10.1103/PhysRevD.92.015001}
  {\path{doi:10.1103/PhysRevD.92.015001}}.

\bibitem{Flacher:2008zq}
H.~Flacher, M.~Goebel, J.~Haller, A.~Hocker, K.~Monig, J.~Stelzer, {Revisiting
  the Global Electroweak Fit of the Standard Model and Beyond with Gfitter},
  Eur. Phys. J. C 60 (2009) 543--583, [Erratum: Eur.Phys.J.C 71, 1718 (2011)].
\newblock \href {http://arxiv.org/abs/0811.0009} {\path{arXiv:0811.0009}},
  \href {https://doi.org/10.1140/epjc/s10052-009-0966-6}
  {\path{doi:10.1140/epjc/s10052-009-0966-6}}.

\bibitem{Baak:2014ora}
M.~Baak, J.~C\'uth, J.~Haller, A.~Hoecker, R.~Kogler, K.~M\"onig, M.~Schott,
  J.~Stelzer, {The global electroweak fit at NNLO and prospects for the LHC and
  ILC}, Eur. Phys. J. C 74 (2014) 3046.
\newblock \href {http://arxiv.org/abs/1407.3792} {\path{arXiv:1407.3792}},
  \href {https://doi.org/10.1140/epjc/s10052-014-3046-5}
  {\path{doi:10.1140/epjc/s10052-014-3046-5}}.

\bibitem{Haller:2018nnx}
J.~Haller, A.~Hoecker, R.~Kogler, K.~M\"onig, T.~Peiffer, J.~Stelzer, {Update
  of the global electroweak fit and constraints on two-Higgs-doublet models},
  Eur. Phys. J. C 78~(8) (2018) 675.
\newblock \href {http://arxiv.org/abs/1803.01853} {\path{arXiv:1803.01853}},
  \href {https://doi.org/10.1140/epjc/s10052-018-6131-3}
  {\path{doi:10.1140/epjc/s10052-018-6131-3}}.

\bibitem{deBlas:2021wap}
J.~de~Blas, M.~Ciuchini, E.~Franco, A.~Goncalves, S.~Mishima, M.~Pierini,
  L.~Reina, L.~Silvestrini, {Global analysis of electroweak data in the
  Standard Model} (12 2021).
\newblock \href {http://arxiv.org/abs/2112.07274} {\path{arXiv:2112.07274}}.

\bibitem{Peskin:1991sw}
M.~E. Peskin, T.~Takeuchi, {Estimation of oblique electroweak corrections},
  Phys. Rev. D 46 (1992) 381--409.
\newblock \href {https://doi.org/10.1103/PhysRevD.46.381}
  {\path{doi:10.1103/PhysRevD.46.381}}.

\bibitem{Peskin:1990zt}
M.~E. Peskin, T.~Takeuchi, {A New constraint on a strongly interacting Higgs
  sector}, Phys. Rev. Lett. 65 (1990) 964--967.
\newblock \href {https://doi.org/10.1103/PhysRevLett.65.964}
  {\path{doi:10.1103/PhysRevLett.65.964}}.

\bibitem{Maksymyk:1993zm}
I.~Maksymyk, C.~P. Burgess, D.~London, {Beyond S, T and U}, Phys. Rev. D 50
  (1994) 529--535.
\newblock \href {http://arxiv.org/abs/hep-ph/9306267}
  {\path{arXiv:hep-ph/9306267}}, \href
  {https://doi.org/10.1103/PhysRevD.50.529}
  {\path{doi:10.1103/PhysRevD.50.529}}.

\bibitem{Buchalla:1995vs}
G.~Buchalla, A.~J. Buras, M.~E. Lautenbacher, {Weak decays beyond leading
  logarithms}, Rev. Mod. Phys. 68 (1996) 1125--1144.
\newblock \href {http://arxiv.org/abs/hep-ph/9512380}
  {\path{arXiv:hep-ph/9512380}}, \href
  {https://doi.org/10.1103/RevModPhys.68.1125}
  {\path{doi:10.1103/RevModPhys.68.1125}}.

\bibitem{Alloul:2013bka}
A.~Alloul, N.~D. Christensen, C.~Degrande, C.~Duhr, B.~Fuks, {FeynRules 2.0 - A
  complete toolbox for tree-level phenomenology}, Comput. Phys. Commun. 185
  (2014) 2250--2300.
\newblock \href {http://arxiv.org/abs/1310.1921} {\path{arXiv:1310.1921}},
  \href {https://doi.org/10.1016/j.cpc.2014.04.012}
  {\path{doi:10.1016/j.cpc.2014.04.012}}.

\bibitem{Hahn:2000kx}
T.~Hahn, {Generating Feynman diagrams and amplitudes with FeynArts 3}, Comput.
  Phys. Commun. 140 (2001) 418--431.
\newblock \href {http://arxiv.org/abs/hep-ph/0012260}
  {\path{arXiv:hep-ph/0012260}}, \href
  {https://doi.org/10.1016/S0010-4655(01)00290-9}
  {\path{doi:10.1016/S0010-4655(01)00290-9}}.

\bibitem{Mertig:1990an}
R.~Mertig, M.~Bohm, A.~Denner, {FEYN CALC: Computer algebraic calculation of
  Feynman amplitudes}, Comput. Phys. Commun. 64 (1991) 345--359.
\newblock \href {https://doi.org/10.1016/0010-4655(91)90130-D}
  {\path{doi:10.1016/0010-4655(91)90130-D}}.

\bibitem{Shtabovenko:2016sxi}
V.~Shtabovenko, R.~Mertig, F.~Orellana, {New Developments in FeynCalc 9.0},
  Comput. Phys. Commun. 207 (2016) 432--444.
\newblock \href {http://arxiv.org/abs/1601.01167} {\path{arXiv:1601.01167}},
  \href {https://doi.org/10.1016/j.cpc.2016.06.008}
  {\path{doi:10.1016/j.cpc.2016.06.008}}.

\bibitem{Shtabovenko:2020gxv}
V.~Shtabovenko, R.~Mertig, F.~Orellana, {FeynCalc 9.3: New features and
  improvements}, Comput. Phys. Commun. 256 (2020) 107478.
\newblock \href {http://arxiv.org/abs/2001.04407} {\path{arXiv:2001.04407}},
  \href {https://doi.org/10.1016/j.cpc.2020.107478}
  {\path{doi:10.1016/j.cpc.2020.107478}}.

\bibitem{Patel:2016fam}
H.~H. Patel, {Package-X 2.0: A Mathematica package for the analytic calculation
  of one-loop integrals}, Comput. Phys. Commun. 218 (2017) 66--70.
\newblock \href {http://arxiv.org/abs/1612.00009} {\path{arXiv:1612.00009}},
  \href {https://doi.org/10.1016/j.cpc.2017.04.015}
  {\path{doi:10.1016/j.cpc.2017.04.015}}.

\bibitem{DeBlas:2019ehy}
J.~De~Blas, et~al., {$\texttt{HEPfit}$: a code for the combination of indirect
  and direct constraints on high energy physics models}, Eur. Phys. J. C 80~(5)
  (2020) 456.
\newblock \href {http://arxiv.org/abs/1910.14012} {\path{arXiv:1910.14012}},
  \href {https://doi.org/10.1140/epjc/s10052-020-7904-z}
  {\path{doi:10.1140/epjc/s10052-020-7904-z}}.

\bibitem{Altmannshofer:2014rta}
W.~Altmannshofer, D.~M. Straub, {New physics in $b\rightarrow s$ transitions
  after LHC run 1}, Eur. Phys. J. C 75~(8) (2015) 382.
\newblock \href {http://arxiv.org/abs/1411.3161} {\path{arXiv:1411.3161}},
  \href {https://doi.org/10.1140/epjc/s10052-015-3602-7}
  {\path{doi:10.1140/epjc/s10052-015-3602-7}}.

\bibitem{Straub:2018kue}
D.~M. Straub, {flavio: a Python package for flavour and precision phenomenology
  in the Standard Model and beyond} (10 2018).
\newblock \href {http://arxiv.org/abs/1810.08132} {\path{arXiv:1810.08132}}.

\bibitem{Li:2021qyo}
X.-Q. Li, M.~Shen, D.-Y. Wang, Y.-D. Yang, X.-B. Yuan, {Explaining the $b \to s
  \ell^+ \ell^-$ anomalies in $Z^\prime$ scenarios with top-FCNC couplings} (12
  2021).
\newblock \href {http://arxiv.org/abs/2112.14215} {\path{arXiv:2112.14215}}.

\bibitem{Geng:2021nhg}
L.-S. Geng, B.~Grinstein, S.~J\"ager, S.-Y. Li, J.~Martin~Camalich, R.-X. Shi,
  {Implications of new evidence for lepton-universality violation in $b \to s
  \ell^+ \ell^-$ decays}, Phys. Rev. D 104~(3) (2021) 035029.
\newblock \href {http://arxiv.org/abs/2103.12738} {\path{arXiv:2103.12738}},
  \href {https://doi.org/10.1103/PhysRevD.104.035029}
  {\path{doi:10.1103/PhysRevD.104.035029}}.

\bibitem{Altmannshofer:2021qrr}
W.~Altmannshofer, P.~Stangl, {New physics in rare B decays after Moriond 2021},
  Eur. Phys. J. C 81~(10) (2021) 952.
\newblock \href {http://arxiv.org/abs/2103.13370} {\path{arXiv:2103.13370}},
  \href {https://doi.org/10.1140/epjc/s10052-021-09725-1}
  {\path{doi:10.1140/epjc/s10052-021-09725-1}}.

\bibitem{Cornella:2021sby}
C.~Cornella, D.~A. Faroughy, J.~Fuentes-Martin, G.~Isidori, M.~Neubert,
  {Reading the footprints of the B-meson flavor anomalies}, JHEP 08 (2021) 050.
\newblock \href {http://arxiv.org/abs/2103.16558} {\path{arXiv:2103.16558}},
  \href {https://doi.org/10.1007/JHEP08(2021)050}
  {\path{doi:10.1007/JHEP08(2021)050}}.

\bibitem{Alguero:2021anc}
M.~Alguer\'o, B.~Capdevila, S.~Descotes-Genon, J.~Matias, M.~Novoa-Brunet,
  {$b\to s\ell\ell$ Global Fits after $R_{K_S}$ and $R_{K^{*+}}$} (4 2021).
\newblock \href {http://arxiv.org/abs/2104.08921} {\path{arXiv:2104.08921}}.

\bibitem{Hurth:2021nsi}
T.~Hurth, F.~Mahmoudi, D.~M. Santos, S.~Neshatpour, {More indications for
  lepton nonuniversality in $b \to s \ell^+ \ell^-$}, Phys. Lett. B 824 (2022)
  136838.
\newblock \href {http://arxiv.org/abs/2104.10058} {\path{arXiv:2104.10058}},
  \href {https://doi.org/10.1016/j.physletb.2021.136838}
  {\path{doi:10.1016/j.physletb.2021.136838}}.

\bibitem{ATLAS:2018ziw}
M.~Aaboud, et~al., {Combination of the searches for pair-produced vector-like
  partners of the third-generation quarks at $\sqrt{s} =$ 13 TeV with the ATLAS
  detector}, Phys. Rev. Lett. 121~(21) (2018) 211801.
\newblock \href {http://arxiv.org/abs/1808.02343} {\path{arXiv:1808.02343}},
  \href {https://doi.org/10.1103/PhysRevLett.121.211801}
  {\path{doi:10.1103/PhysRevLett.121.211801}}.

\bibitem{CMS:2022yxp}
A.~Tumasyan, et~al., {Search for single production of a vector-like T quark
  decaying to a top quark and a Z boson in the final state with jets and
  missing transverse momentum at $\sqrt{s}$ = 13 TeV} (1 2022).
\newblock \href {http://arxiv.org/abs/2201.02227} {\path{arXiv:2201.02227}}.

\bibitem{Aguilar-Saavedra:2013qpa}
J.~A. Aguilar-Saavedra, R.~Benbrik, S.~Heinemeyer, M.~P\'erez-Victoria,
  {Handbook of vectorlike quarks: Mixing and single production}, Phys. Rev. D
  88~(9) (2013) 094010.
\newblock \href {http://arxiv.org/abs/1306.0572} {\path{arXiv:1306.0572}},
  \href {https://doi.org/10.1103/PhysRevD.88.094010}
  {\path{doi:10.1103/PhysRevD.88.094010}}.

\bibitem{delaTorreTrishaFarooque:2022vqc}
H.~de~la TorreTrisha~Farooque, T.~Farooque, {Looking beyond the Standard Model
  with Third Generation Quarks at the LHC}, Symmetry 14~(3) (2022) 444.
\newblock \href {https://doi.org/10.3390/sym14030444}
  {\path{doi:10.3390/sym14030444}}.

\bibitem{Serra:2015xfa}
J.~Serra, {Beyond the Minimal Top Partner Decay}, JHEP 09 (2015) 176.
\newblock \href {http://arxiv.org/abs/1506.05110} {\path{arXiv:1506.05110}},
  \href {https://doi.org/10.1007/JHEP09(2015)176}
  {\path{doi:10.1007/JHEP09(2015)176}}.

\bibitem{Anandakrishnan:2015yfa}
A.~Anandakrishnan, J.~H. Collins, M.~Farina, E.~Kuflik, M.~Perelstein, {Odd Top
  Partners at the LHC}, Phys. Rev. D 93~(7) (2016) 075009.
\newblock \href {http://arxiv.org/abs/1506.05130} {\path{arXiv:1506.05130}},
  \href {https://doi.org/10.1103/PhysRevD.93.075009}
  {\path{doi:10.1103/PhysRevD.93.075009}}.

\bibitem{Bizot:2018tds}
N.~Bizot, G.~Cacciapaglia, T.~Flacke, {Common exotic decays of top partners},
  JHEP 06 (2018) 065.
\newblock \href {http://arxiv.org/abs/1803.00021} {\path{arXiv:1803.00021}},
  \href {https://doi.org/10.1007/JHEP06(2018)065}
  {\path{doi:10.1007/JHEP06(2018)065}}.

\bibitem{ATLAS:2019erb}
G.~Aad, et~al., {Search for high-mass dilepton resonances using 139 fb$^{-1}$
  of $pp$ collision data collected at $\sqrt{s}=$13 TeV with the ATLAS
  detector}, Phys. Lett. B 796 (2019) 68--87.
\newblock \href {http://arxiv.org/abs/1903.06248} {\path{arXiv:1903.06248}},
  \href {https://doi.org/10.1016/j.physletb.2019.07.016}
  {\path{doi:10.1016/j.physletb.2019.07.016}}.

\bibitem{CMS:2021ctt}
A.~M. Sirunyan, et~al., {Search for resonant and nonresonant new phenomena in
  high-mass dilepton final states at $ \sqrt{s} $ = 13 TeV}, JHEP 07 (2021)
  208.
\newblock \href {http://arxiv.org/abs/2103.02708} {\path{arXiv:2103.02708}},
  \href {https://doi.org/10.1007/JHEP07(2021)208}
  {\path{doi:10.1007/JHEP07(2021)208}}.

\bibitem{Borah:2022zim}
D.~Borah, S.~Mahapatra, N.~Sahu, {Singlet-Doublet Fermion Origin of Dark
  Matter, Neutrino Mass and W-Mass Anomaly} (4 2022).
\newblock \href {http://arxiv.org/abs/2204.09671} {\path{arXiv:2204.09671}}.

\end{thebibliography}

\end{document}